# Removing the divergence of Chandrasekhar limit caused by generalized uncertainty principle

Xin-Dong Du[1], Chao-Yun Long[2]

Department of Physics, Guizhou University, Guiyang 550025, China

**Abstract**

The usual generalized uncertainty principle will lead to a divergent mass limit of white dwarf, and this divergence should be prevented for both scenarios including positive and negative parameters of generalized uncertainty principle. Although it has been shown that negative parameter can directly restore the mass limit, the underlying reason is not given to explain why the negative sign appears under the condition of white dwarf. In order to solve this problem, we derive a field-dependent parameter expression whose sign can change depending on the species of spin fields. Besides, we find that the actual physical effect of the negative sign is aimed at limiting the exorbitant uncertainty of momentum.

## 1  Introduction

White dwarfs are dense stars and their electrons are completely degenerate to produce an outward push, known as degenerate electron pressure [1], to support the stars. This situation roots in two principles that govern the very small quantum state: Heisenberg uncertainty principle [2] and Pauli exclusion principle [3]. A stable white dwarf needs its gravitational potential energy to balance its kinetic energy of the electrons, namely:

$$|E_g| \leq |E_k|, \qquad (1)$$

where $E_g$ is the gravitational potential energy of white dwarf, and $E_k$ is the maximum kinetic energy that its electrons can provide. The specific expressions of $E_g$ and $E_k$ are as follows:

$$E_g = -\frac{3}{5}\frac{GM^2}{R}, \qquad (2)$$

$$E_k = P_e\left(\frac{4}{3}\pi R^3\right), \qquad (3)$$

where $P_e$ is the degenerate electron pressure, and its expression for the non-relativistic

---
[1] First author.
[2] Corresponding author. E-mail: LongCYGuiZhou@163.com

case is given by [1]:

$$P_e \approx \left(\frac{3}{8\pi}\right)^{\frac{2}{3}} \frac{h^2}{5 m_e m_p^{\frac{5}{3}}} \left(\frac{Z}{A}\right)^{\frac{5}{3}} \rho^{\frac{5}{3}}, \qquad (4)$$

where $h$ is the Planck constant, $m_e$ is the electron mass, $m_p$ is the proton mass, $Z$ is the atomic number, $A$ is the atomic mass number, and $\rho$ is the mass density.

However, $E_k$ is the maximum value that the electrons can provide, so it should be gained from electrons in highly compressed state. It means the speed of degenerate matter will reach the order of the speed of light, and the special relativity must be considered. In [4], the degenerate electron pressure for the relativistic case is yielded:

$$P_e \approx \left(\frac{3}{8\pi}\right)^{\frac{1}{3}} \frac{hc}{4 m_p^{\frac{4}{3}}} \left(\frac{Z}{A}\right)^{\frac{4}{3}} \rho^{\frac{4}{3}}, \qquad (5)$$

from which we can see the electron mass does not appear and the degenerate electron pressure is completely dominated by the proton mass. The more detailed equation of state of degenerate electron gas in the extreme relativistic limit is derived in [5], and the critical mass of white dwarf becomes:

$$M_c = 0.21 \left(\frac{Z}{A}\right)^2 \left(\frac{hc}{G}\right)^{\frac{3}{2}} \frac{1}{m_p^2}, \qquad (6)$$

which is known as Chandrasekhar limit and agrees well with most observations [6-9]. For white dwarf, the main elements are helium, carbon and oxygen, so $Z/A \approx 0.5$. This leads to:

$$M_c \approx 1.46 M_\odot, \qquad (7)$$

where the solar mass $M_\odot = 1.989 \times 10^{30}$ kg.

String theory [10], Gedanken experiment [11], non-commutativity [12] and other models of quantum gravity [13] indicate that the Heisenberg uncertainty principle should be modified to reflect the existence of a minimal length scale. As a result, the generalized uncertainty principle is proposed [14-19]. However, the corrected mass limit of white dwarf from the generalized uncertainty principle will diverge and exceed Chandrasekhar limit [20], while the Heisenberg uncertainty principle does not cause this divergence. It means that some necessary solutions need to be given for the generalized uncertainty principle. The parameter signs of generalized uncertainty principles would change with different quantum gravity models [21, 22], and positive parameter leads to a minimum length while negative one removes the minimum length

(negative parameter lets $\Delta x \Delta p \geq 0$, so physics becomes classical and there is no minimum length) [22]. Moreover, these behaviors of parameter signs can be clearly reflected in the end state of Hawking evaporation: positive parameter gives rise to a black hole remnant [23], and negative parameter would not result in an actual remnant [24]. For these reasons, we are going to prevent this divergence for both scenarios including positive and negative parameters. It should be pointed out that parameter signs are not limited to the usual generalized uncertainty principle but also include the improved. Although improved generalized uncertainty principles [25-27] own different forms from the usual one, most of them can still be reduced to the usual form by omitting some high-order terms or additional terms. In other words, the parameter signs should have the same effect on them: positive sign stands for including minimum length and negative sign stands for excluding minimum length. And in order to show the effect of parameter signs uniformly, we will use a common notation $\alpha$ to represent these parameters in our paper.

When parameter is positive, a high-momentum limit is provided by a corrected Tolman–Oppenheimer–Volkoff equations in [28], which restores the mass limit for the usual generalized uncertainty principle. Therefore, the case of positive parameter has been resolved well and we are going to focus on the other case that involves negative parameter. When parameter is negative, it has been shown in [29] that the divergence can be removed by a negative sign, but the underlying reason is not given to explain why the negative sign appears under the condition of white dwarf. In our paper, we aim to derive a field-dependent parameter expression whose sign can change depending on the species of spin fields, and the reason for the occurrence of negative sign is expected to be explained by this expression. Besides, in order to explore a connection between both scenarios, we will also offer a solution for the case of positive parameter.

The mass-radius relations of white dwarf for the relativistic case are usually derived to describe the mass limit [29, 30]. However, their results show that the mass is sometimes positively correlated with the radius disagreeing with most observations [6-8]. The reason for the unreasonable results is that the used mass-radius relations reflecting the mass limit are only suitable for the relativistic case and should not be extended to non-relativistic case. If a white dwarf does not approach its mass limit, then the relativistic case will not be set up and the original mass-radius relation should be inapplicable. But a complete mass-radius relation is too hard to gain because we do not know the exact range of mass for the relativistic case. Thus, our paper will use inequalities instead of a mass-radius relation to judge whether the white dwarf mass limit diverges. By applying inequalities, the final conclusions can be directly reflected in the directions of inequality signs, which is much simpler and clearer.

This paper is organized as follows: in Sect. 2, we introduce an approximation method to calculate the Chandrasekhar limit from uncertainty principles. In Sect. 3, the approximation method is used to reproduce the divergent mass limit of white dwarf from the generalized uncertainty principle. In Sect. 4, when parameter is positive, we use an improved generalized uncertainty principle with maximum momentum to remove the divergence. In Sect. 5, when parameter is negative, a field-dependent parameter expression for the generalized uncertainty principle is derived. The expression not only explains the occurrence of negative sign that restores the mass limit, but also lets parameter be positive in other cases to be consistent with string theory and some models of quantum gravity. Also, the rationality of the expression is discussed. In Sect. 6, we give the actual physical effect of the negative sign and analyze the internal connection between the solutions in Sect. 4 and Sect. 5.

## 2  Approximation method to calculate mass limit

Generally, the mathematical derivation of the Chandrasekhar limit requires complex differentials and integrals [20], which lets it hard to see clearly how uncertainty principles influence the Chandrasekhar limit. For this reason, we introduce an approximation method to calculate the limit (see [29] for similar treatment). According to [31], the approximate expression of relativistic degenerate electron pressure $P_e$ is written as:

$$P_e \approx \frac{2\Delta pc}{(\Delta x)^3}, \tag{8}$$

where $(\Delta x)^3$ can be expressed by the white dwarf mass density $\rho$ [31]:

$$(\Delta x)^3 \approx \frac{2V}{N_e} = \frac{2V}{\frac{ZM}{Am_p}} = \frac{2Am_p}{Z\rho}. \tag{9}$$

Eq. (8) and Eq. (9) are the keys to relate uncertainty principles and the Chandrasekhar limit. In order to test their rationality, we are going to use them to get the corresponding Chandrasekhar limit from the Heisenberg uncertainty principle. The Heisenberg uncertainty principle is as follows:

$$\Delta x \Delta p \geq \frac{\hbar}{2}, \tag{10}$$

leading to:

$$\Delta p \approx \frac{\hbar}{2\Delta x}. \tag{11}$$

Eq. (8) can be updated by Eq. (9) and Eq. (11) as:

$$P_e \approx (0.5)^{\frac{4}{3}} \frac{\hbar c}{m_p^{\frac{4}{3}}} \left(\frac{Z}{A}\right)^{\frac{4}{3}} \rho^{\frac{4}{3}}. \tag{12}$$

Comparing Eq. (12) and Eq. (5), we can see that they differ only in the dimensionless constant, which shows the approximation fits well with the theory. The white dwarf mass density $\rho$ can be obtained by:

$$\rho = \frac{M}{\frac{4}{3}\pi R^3}, \tag{13}$$

and therefore:

$$P_e \approx (0.5)^{\frac{4}{3}} \frac{\hbar c}{m_p^{\frac{4}{3}}} \left(\frac{Z}{A}\right)^{\frac{4}{3}} \left(\frac{M}{\frac{4}{3}\pi R^3}\right)^{\frac{4}{3}}. \tag{14}$$

Eq. (1) immediately conveys:

$$\frac{3}{5}\frac{GM^2}{R} \leq (0.5)^{\frac{4}{3}} \frac{\hbar c}{m_p^{\frac{4}{3}}} \left(\frac{Z}{A}\right)^{\frac{4}{3}} \left(\frac{M}{\frac{4}{3}\pi R^3}\right)^{\frac{4}{3}} \left(\frac{4}{3}\pi R^3\right), \tag{15}$$

from which we can get:

$$M_{max} \approx 0.538 \left(\frac{Z}{A}\right)^2 \left(\frac{\hbar c}{G}\right)^{\frac{3}{2}} \frac{1}{m_p^2}, \tag{16}$$

which is the Chandrasekhar limit obtained from the Heisenberg uncertainty principle. And the only difference between Eq. (16) and Eq. (6) is in the dimensionless constant. Thus, we can consider Eq. (8) and Eq. (9) as a reasonable approximation method to calculate the Chandrasekhar limit. In the next section, we will see the approximation method can be used to reproduce the divergent mass limit of white dwarf from the generalized uncertainty principle.

## 3 Divergence from generalized uncertainty principle

Although in [20], it has been shown how to strictly calculate the divergent mass limit of white dwarf from the generalized uncertainty principle, the approximate method provided in Sect. 2 should also be used to reproduce the divergence so as to be compared with the results in Sect. 4 and Sect. 5. The form of the usual generalized uncertainty principle is given by [19]:

$$\Delta x \Delta p \geq \frac{\hbar}{2}\left(1 + \frac{\alpha L_p^2}{\hbar^2}(\Delta p)^2\right), \tag{17}$$

where $\alpha$ is a dimensionless constant that is generally considered to be positive, and $L_p = \sqrt{G\hbar/c^3}$ is the Planck length. We transform Eq. (17) into a perfect square expression of $\Delta p$ and take its square root to obtain:

$$\frac{\Delta x \hbar}{\alpha L_p^2}\left[1 + \sqrt{1 - \frac{\alpha L_p^2}{(\Delta x)^2}}\right] \geq \Delta p \geq \frac{\Delta x \hbar}{\alpha L_p^2}\left[1 - \sqrt{1 - \frac{\alpha L_p^2}{(\Delta x)^2}}\right], \tag{18}$$

leading to:

$$\Delta p \approx \frac{\Delta x \hbar}{\alpha L_p^2}\left[1 - \sqrt{1 - \frac{\alpha L_p^2}{(\Delta x)^2}}\right]. \tag{19}$$

Here we only consider the right-hand part of Eq. (18) because this part can be reduced into the Heisenberg uncertainty principle as $\alpha \to 0$. It should be noted that the above approximation might have a huge impact on deriving precise thermodynamic relations [32]. However, the approximation is going to be used in inequalities to determine the directions of inequality signs, so the final results will not be significantly affected by the approximation.

Eq. (8) can be updated by Eq. (19) as:

$$P_e \approx \frac{2c}{(\Delta x)^2}\frac{\hbar}{\alpha L_p^2}\left[1 - \sqrt{1 - \frac{\alpha L_p^2}{(\Delta x)^2}}\right]. \tag{20}$$

We rewrite the above equation with Eq. (9) and Eq. (13), yielding:

$$P_e \approx \frac{c}{\left(\frac{Am_p}{ZM}\right)^{\frac{2}{3}} R^2}\frac{\hbar}{\alpha L_p^2}\left[1 - \sqrt{1 - \frac{\alpha L_p^2}{\left(\frac{Am_p}{ZM}\right)^{\frac{2}{3}} R^2}}\right]. \tag{21}$$

From the above equation onwards, we omit the dimensionless coefficients except $\alpha$ for simplicity of calculation, and Eq. (1) immediately conveys:

$$\frac{GM^2}{R} \leq \frac{c}{\left(\frac{Am_p}{ZM}\right)^{\frac{2}{3}} R^2} \frac{\hbar}{\alpha L_p^2} \left[1 - \sqrt{1 - \frac{\alpha L_p^2}{\left(\frac{Am_p}{ZM}\right)^{\frac{2}{3}} R^2}}\right] R^3. \tag{22}$$

In fact, the gravitational potential energy $|E_g| = GM^2/R$ should be modified by the generalized uncertainty principle as $|E_g|[1 + (\alpha/144) 5 \ln(4\pi R^2)/(3R^2)]$ [33]. However, the corrected result will not affect our judgement about whether the mass limit diverges, because it only changes the coefficient of $M^2$ instead of the structure of $M^2$. Thus, we will neglect this correction from the generalized uncertainty principle to keep a much simpler and clearer procedure. Let us simplify Eq. (22) as follows:

$$\frac{GM^{\frac{4}{3}}}{cR^2} \frac{\alpha L_p^2}{\hbar} \left(\frac{Am_p}{Z}\right)^{\frac{2}{3}} \leq 1 - \sqrt{1 - \frac{\alpha L_p^2}{\left(\frac{Am_p}{ZM}\right)^{\frac{2}{3}} R^2}}, \tag{23}$$

$$\left(\frac{GM^{\frac{4}{3}}}{cR^2} \frac{\alpha L_p^2}{\hbar} \left(\frac{Am_p}{Z}\right)^{\frac{2}{3}} - 1\right)^2 \geq 1 - \frac{\alpha L_p^2}{\left(\frac{Am_p}{ZM}\right)^{\frac{2}{3}} R^2}, \tag{24}$$

$$\left(\frac{G}{cR^2} \frac{\alpha L_p^2}{\hbar}\right)^2 \left(\frac{Am_p}{Z}\right)^{\frac{4}{3}} M^2 \geq 2 \left(\frac{G}{cR^2} \frac{\alpha L_p^2}{\hbar}\right) \left(\frac{Am_p}{Z}\right)^{\frac{2}{3}} M^{\frac{2}{3}} - \frac{\alpha L_p^2}{\left(\frac{Am_p}{Z}\right)^{\frac{2}{3}} R^2}. \tag{25}$$

It is worth noting that the inequality sign of Eq. (23) should change if $\alpha < 0$. Here we keep $\alpha > 0$ because an effective generalized uncertainty principle needs a positive parameter to reflect minimum length [22]. For large enough $M$, the constant term of Eq. (25) is negligible and we have:

$$\left(\frac{G}{cR^2} \frac{\alpha L_p^2}{\hbar}\right)^2 \left(\frac{Am_p}{Z}\right)^{\frac{4}{3}} M^2 \geq 2 \left(\frac{G}{cR^2} \frac{\alpha L_p^2}{\hbar}\right) \left(\frac{Am_p}{Z}\right)^{\frac{2}{3}} M^{\frac{2}{3}}, \tag{26}$$

leading to:

$$M \geq 2 \left(\frac{Z}{Am_p}\right)^{\frac{1}{2}} \left(\frac{c\hbar R^2}{\alpha G L_p^2}\right)^{\frac{3}{4}}, \tag{27}$$

which indicates that the mass limit of white dwarf corrected by the generalized uncertainty principle is divergent, agreeing with the result in [20] derived by a rigorous calculation. There have been a few observations of type Ia supernovae arguing that their progenitor mass will exceed the Chandrasekhar limit [34, 35]. A possible origin is explained by [36] based on non-commutative geometry [12], and a more thorough explanation is obtained by combining the idea of a squashed fuzzy sphere [37]. However, their masses are still finite and of the same order of magnitude as $M_c$. By contrast, the result from Eq. (27) that allows the mass limit to diverge to infinity, is so disconcerting that we need to improve it.

Because different quantum gravity models would correspond to different parameter signs of generalized uncertainty principles [21, 22], the divergence of mass limit should be respectively removed from both scenarios including positive and negative parameters. The case of positive parameter has been resolved well by [28], so our paper will focus on the case of negative parameter. However, in order to explore a connection between the two cases, we first need to provide a definite limit for momentum in the case of positive parameter, but the solution in [28] gives a similar limit about electroweak bound instead of momentum. Thus, for completeness of analysis, an improved generalized uncertainty principle with maximum momentum will be used in the next section to restore the mass limit in the case of positive parameter. And the specific analysis for the connection between the two cases will be shown in Sect. 6.

## 4  Removing the divergence when parameter is positive

In this section, we aim to provide a definite limit for momentum to remove the divergence in the case of positive parameter, which can be compared more directly with the result of negative parameter so as to allow us to explore a connection between the two cases. In doubly special relativity theories [38], the existence of a maximal momentum is essentially needed to keep velocity of light and the Planck energy invariable [25], so the maximal momentum should be included in the modification of generalized uncertainty principle. An improved generalized uncertainty principle is proposed by [25], which is consistent with doubly special relativity and predicts both a minimum length and a maximum momentum:

$$\Delta x \Delta p \geq \frac{\hbar}{2}\left[1 + \left(\frac{\alpha}{\sqrt{\langle p^2 \rangle}} + 4\alpha^2\right)(\Delta p)^2 + 4\alpha^2 \langle p \rangle^2 - 2\alpha\sqrt{\langle p^2 \rangle}\right], \quad (28)$$

where both a minimum length and a maximum momentum are given at the same time:

$$\begin{cases} \Delta x \geq (\Delta x)_{min} \approx \alpha_0 L_p \\ \Delta p \leq (\Delta p)_{max} \approx \dfrac{M_p c}{\alpha_0} \end{cases}, \tag{29}$$

where $\alpha = \alpha_0/(M_p c) = \alpha_0 L_p / \hbar > 0$, $\alpha_0 > 0$ is a dimensionless parameter and $M_p = \sqrt{\hbar c / G}$ is the Planck mass.

Eq. (8) is updated by Eq. (29) as:

$$(P_e)_{max} \approx \frac{2(\Delta p)_{max} c}{(\Delta x)^3} \approx \frac{2 M_p c^2}{\alpha_0 (\Delta x)^3}. \tag{30}$$

The above equation can be deformed with Eq. (9) and Eq. (13):

$$(P_e)_{max} \approx \frac{3}{4\pi} \frac{M_p c^2}{\alpha_0} \left(\frac{ZM}{A m_p}\right) \frac{1}{R^3}. \tag{31}$$

According to Eq. (1), we have:

$$\frac{3}{5} \frac{G M^2}{R} \leq P_e \left(\frac{4}{3} \pi R^3\right) \leq (P_e)_{max} \left(\frac{4}{3} \pi R^3\right), \tag{32}$$

where we have neglected the correction of the gravitational potential energy from the improved generalized uncertainty principle. Because Eq. (28) has a similar main form with Eq. (17), the corrected result from the improved one will not affect our judgement about whether the mass limit diverges, as discussed in Sect. 3. We can update Eq. (32) by Eq. (31) as:

$$\frac{3}{5} \frac{G M^2}{R} \leq \frac{3}{4\pi} \frac{M_p c^2}{\alpha_0} \left(\frac{ZM}{A m_p}\right) \frac{1}{R^3} \left(\frac{4}{3} \pi R^3\right), \tag{33}$$

leading to:

$$M \leq \frac{5}{3\alpha_0} \left(\frac{Z}{A}\right) \left(\frac{M_p}{m_p}\right) \left(\frac{R c^2}{G}\right). \tag{34}$$

The only variable in Eq. (34) is the radius $R$ of white dwarf, and it must be finite because $R$ is inversely correlated with $M$ [6-8, 39]. Thus, if white dwarf can keep stable under gravity, its mass should be less than the finite value provided by Eq. (34). In other words, the mass limit can be restored when a maximum momentum is embodied in the generalized uncertainty principle according to Eq. (28).

The divergence problem in the case of positive parameter has been resolved in [28] by incorporating the usual generalized uncertainty principle. In this section, we explore another solution for the case of positive parameter from an improved generalized uncertainty principle, and the difference from the former is that a definite limit for momentum is obtained. Besides, a similar investigation is shown in [40], which gets a maximum momentum from a higher order generalized uncertainty principle. In the next section, we will see that the mass limit can also be restored by a negative sign of parameter. In order to find the internal connection between both the solutions under different parameter signs, we will combine these results given in this section and the next section for further analysis (see Sect. 6).

## 5 Removing the divergence when parameter is negative

Although the case of negative parameter has been discussed in [29], the underlying reason is not given to explain why the negative sign appears under the condition of white dwarf. In this section, we aim to derive a field-dependent parameter expression to solve the problem completely. To begin with, we need to get a specific expression of the parameter. In [41], the Schwarzschild metric is connected with the generalized uncertainty principle by the Hawking temperature to give an analytic relation between the deformation parameter of metric and the parameter of generalized uncertainty principle. Following this idea, we are going to connect the quantum tunneling [42] with the usual generalized uncertainty principle by the logarithmic term [43] of the Bekenstein-Hawking entropy, so as to obtain a relation between the trace anomaly [44, 45] and the parameter of the usual generalized uncertainty principle. In order to make the result more general, the Bekenstein-Hawking entropy will be gained from Kerr-Newman black hole. From here onwards, $k_B = c = \hbar = G = 1$ will be set.

According to [46], the Bekenstein-Hawking entropy of Kerr-Newman black hole can be corrected by the usual generalized uncertainty principle, namely:

$$S_{BH} = \frac{\pi}{2}\left[\rho_0{}^2 + \rho_0\sqrt{\rho_0{}^2 - \alpha} - \alpha \ln\left(\rho_0 + \sqrt{\rho_0{}^2 - \alpha}\right)\right], \qquad (35)$$

where $\alpha$ is the parameter of the usual generalized uncertainty principle and the black hole area $A_{BH} = 4\pi\rho_0{}^2$. When $\rho_0{}^2 \gg \alpha$ as is assumed in [46], we have:

$$S_{BH} \simeq \frac{A_{BH}}{4} - \frac{\pi\alpha}{4}\ln(A_{BH}) + \frac{\pi\alpha}{4}\ln(\pi). \qquad (36)$$

$S_{BH}$ can also be derived in [47] by quantum tunneling:

$$S_{BH} \simeq \frac{A_{BH}}{4} + 2\pi\alpha_1 \ln(A_{BH}) + \text{const} + \text{higher order terms}, \qquad (37)$$

where $\alpha_1$ is provided by:

$$\alpha_1 = -\frac{\left(M^2 - Q^2 - \frac{J^2}{M^2}\right)^{\frac{1}{2}}}{4\pi\omega} \text{Im} \int d^4x \sqrt{-g}\, \langle T^\mu_\mu \rangle. \qquad (38)$$

Here the integral part of Eq. (38) is the integral of the trace anomaly.

On the one hand, both quantum tunneling and generalized uncertainty principle are the theoretical models to reflect the existence of minimum length [48, 49], so the results from them should be equivalent to each other. On the other hand, the logarithmic term of black hole entropy is regarded as the leading correction, and it is widely accepted in [43, 50, 51]. Hence, we can directly link the logarithmic terms of Eq. (36) and Eq. (37) together, and the relation between $\alpha$ and $\alpha_1$ will become:

$$\alpha = -8\alpha_1 = \frac{2}{\pi}\frac{\left(M^2 - Q^2 - \frac{J^2}{M^2}\right)^{\frac{1}{2}}}{\omega} \text{Im} \int d^4x \sqrt{-g} \langle T^\mu_\mu \rangle. \tag{39}$$

Using $\omega = \left(M^2 - Q^2 - \frac{J^2}{M^2}\right)^{\frac{1}{2}}$ provided in [47] for Kerr-Newman black hole, we can turn Eq. (39) into:

$$\alpha = \frac{2}{\pi} \text{Im} \int d^4x \sqrt{-g} \langle T^\mu_\mu \rangle. \tag{40}$$

In [52, 53], another expression of the integral of the trace anomaly is given by:

$$\text{Im} \int d^4x \sqrt{-g} \langle T^\mu_\mu \rangle = \frac{1}{45}\left(-N_0 - \frac{7}{4}N_{\frac{1}{2}} + 13N_1 + \frac{233}{4}N_{\frac{3}{2}} - 212N_2\right), \tag{41}$$

where $N_s$ is the number of fields with spin 's'. Combining Eq. (40) and Eq. (41), we can express the parameter of the usual generalized uncertainty principle as a function of $N_s$:

$$\alpha = \frac{2}{45\pi}\left(-N_0 - \frac{7}{4}N_{\frac{1}{2}} + 13N_1 + \frac{233}{4}N_{\frac{3}{2}} - 212N_2\right). \tag{42}$$

In string theory, the uncertainty principle follows as a direct consequence of the quantization of electromagnetic radiation in the form of photons [54] and the spin of photons is 1, leading to $\alpha = 26/(45\pi) N_1 > 0$. It means the parameter expression Eq. (42) reflects a widely accepted positive parameter of the usual generalized uncertainty principle for string theory and some models of quantum gravity [15, 54].

In the case of white dwarf, the generalized uncertainty principle will be used to calculate the electron pressure in extreme degenerate case, so $N_s$ should stand for the distribution of fields around the electrons of white dwarf. The electron gas in extreme degenerate case can be regard as the ideal Fermi gas where the particles are not interacting [55], allowing the spin 1 / 2 of electrons to become dominant (here we have ignored the effects of other particles on the spin of electrons). Thus, the parameter $\alpha$ in this case can be expressed as:

$$\alpha = -\frac{7}{90\pi} N_{\frac{1}{2}} < 0, \tag{43}$$

which shows that the parameter of the usual generalized uncertainty principle should be negative under the condition of white dwarf. The parameter expression given by Eq. (42) offers a good explanation for the appearance of the negative parameter, and this point is not deeply studied in [29]. Again, it is necessary to emphasize: the negative parameter makes $\Delta x \Delta p \geq 0$, leading to a classical physics around the Planck scale [22], so there is no minimum length. For this reason, the minimum length should be regarded as being excluded when we use the negative parameter. Taking the negative parameter into the calculation of the mass limit of white dwarf, we can see the procedures are almost the same as that in Sect. 3, and the only difference is that the directions of inequality signs of Eq. (23) and every inequality after it need to be reversed:

$$\frac{GM^{\frac{4}{3}}}{cR^2} \frac{\alpha L_p^2}{\hbar} \left(\frac{Am_p}{Z}\right)^{\frac{2}{3}} \geq 1 - \sqrt{1 - \frac{\alpha L_p^2}{\left(\frac{Am_p}{ZM}\right)^{\frac{2}{3}} R^2}}, \tag{44}$$

$$\left(\frac{GM^{\frac{4}{3}}}{cR^2} \frac{\alpha L_p^2}{\hbar} \left(\frac{Am_p}{Z}\right)^{\frac{2}{3}} - 1\right)^2 \leq 1 - \frac{\alpha L_p^2}{\left(\frac{Am_p}{ZM}\right)^{\frac{2}{3}} R^2}, \tag{45}$$

$$\left(\frac{G}{cR^2} \frac{\alpha L_p^2}{\hbar}\right)^2 \left(\frac{Am_p}{Z}\right)^{\frac{4}{3}} M^2 \leq 2 \left(\frac{G}{cR^2} \frac{\alpha L_p^2}{\hbar}\right) \left(\frac{Am_p}{Z}\right)^{\frac{2}{3}} M^{\frac{2}{3}} - \frac{\alpha L_p^2}{\left(\frac{Am_p}{Z}\right)^{\frac{2}{3}} R^2}. \tag{46}$$

Let us remove the negative term of the right-hand part of Eq. (46), and then its inequality sign will not be altered:

$$\left(\frac{G}{cR^2} \frac{\alpha L_p^2}{\hbar}\right)^2 \left(\frac{Am_p}{Z}\right)^{\frac{4}{3}} M^2 \leq -\frac{\alpha L_p^2}{\left(\frac{Am_p}{Z}\right)^{\frac{2}{3}} R^2}, \tag{47}$$

leading to:

$$M \leq \frac{1}{\sqrt{-\alpha}} \left(\frac{Z}{A}\right) \left(\frac{R}{L_p}\right) \left(\frac{\hbar c}{G}\right) \frac{1}{m_p}, \tag{48}$$

which indicates that the mass limit can be restored when the field-dependent parameter expression Eq. (42) is used for white dwarf to provide a negative sign. It should be noted: although the field-dependent parameter expression gives a negative parameter for white dwarf, we never deny the case of positive parameter shown in Sect. 4. In fact, different quantum gravity models will give rise to different parameter signs [21, 22], so we should just regard them as two separate scenarios. And because the divergence can be prevented for both scenarios, we need not subscribe to either of these scenarios.

String theory and some models of quantum gravity give rise to a positive parameter [15, 54], while a negative parameter could be obtained in the case of white dwarf. Hence, we need to explain the essential reason for the change of parameter signs, and the field-dependent parameter expression Eq. (42) makes it possible. On the one hand, string theory reflects the quantization of electromagnetic radiation in the form of photons [54], so fields with spin 1 become dominant, leading to a positive sign. On the other hand, the electron gas of white dwarf in extreme degenerate case can be regard as the ideal Fermi gas [55], so fields with spin $1/2$ become dominant, leading to a negative sign. Besides, another improved generalized uncertainty principle including cosmological constant is introduced by [30] to restore the mass limit, and it allows a wide range of values of both positive and negative parameter. This new attempt in [30] might provide another suitable path to further explore the conversion between parameter signs.

In our paper, the parameter of the usual generalized uncertainty principle is not a constant but a function of the number of spin fields. The parameter forms have been studied in many articles. Some papers [22, 56-58] give specific numerical values for the parameter by comparing the results obtained in two independent models. However, they get different specific values depending on the different used models, which indicates the parameter is more likely to be a function rather than a fixed value. A functional expression for the parameter based on mass is gained in [59], but the result is from Schwarzschild black hole and the object for comparison is not a generally accepted leading term. By contrast, we obtain a field-dependent parameter expression from a more general Kerr-Newman black hole, and the object for comparison is the widely accepted logarithmic term of entropy. In addition, it can be seen from Eq. (42) that the parameter is related to the number of spin fields, agreeing with the prediction in [60]. Taking these aspects together, we argue that the obtained field-dependent parameter expression is reasonable and persuasive.

## 6 Discussions and conclusions

We have shown two independent solutions to restore the mass limit of white dwarf from two scenarios including positive and negative parameters. For the case of positive parameter, we can see the divergence can be removed by a maximal momentum, agreeing with the suggestions given in [28, 40]. But for the case of negative parameter, there is still a lack of physical meaning to explain why the negative sign can restore the mass limit. Now we are going to give a specific physical meaning about the negative sign, and the negative parameter $\alpha$ can turn the generalized uncertainty principle Eq. (17) into:

$$\Delta x \geq \frac{\hbar}{2} f(\Delta p), \tag{49}$$

where $f(\Delta p)$ is:

$$f(\Delta p) = \frac{1}{\Delta p} - \frac{|\alpha| L_p^2}{\hbar^2} \Delta p. \tag{50}$$

It is obvious that Eq. (49) would be meaningless if $f(\Delta p) < 0$, so we must keep $f(\Delta p) \geq 0$ to ensure the generalized uncertainty principle is essentially reasonable. And because Eq. (50) is a decreasing function of $\Delta p$, we should ensure:

$$\Delta p \leq \frac{\hbar}{\sqrt{|\alpha|} L_p} = \frac{M_p c}{\sqrt{|\alpha|}}. \tag{51}$$

The above inequality shows that the negative sign provides a severe restriction on the allowed values of momentum, which is consistent with the suggestion from the case of positive parameter in Sect. 4. In fact, according to Eq. (30) and Eq. (32), we can see that the limit for momentum leads to a limit for degenerate electron pressure and the size of degenerate electron pressure determines the size of white dwarf maximum mass. Hence, the negative sign can successfully restore the mass limit by providing a limit for momentum.

Eq. (29) and Eq. (51) are respectively the results from the cases of positive parameter and negative parameter. Comparing Eq. (29) and Eq. (51), we can see that they are exactly the same in form except for dimensionless coefficient. It further indicates that the solutions in the two cases serve a common purpose: to prevent the uncertainty of momentum from being too large. It is interesting to explore why the common purpose needs to be provided to restore the mass limit. Considering that the Heisenberg uncertainty principle can keep the mass limit convergent without limiting the momentum, the common purpose should arise from the structural defect of the usual generalized uncertainty principle. For example, the existence of $(\Delta p)^2$ in Eq. (17) might need an additional limit for momentum to maintain its rationality. We hope this problem could be comprehensively solved in the future.

In conclusion, we first give an approximation method to calculate the mass limit of white dwarf, so we can analyze how the uncertainty principles influence the Chandrasekhar limit with a clearer view. And we prove the approximation method is feasible by comparing its results with the relativistic degenerate electron pressure and the Chandrasekhar limit. After that, we apply the method to reproduce the divergent mass limit of white dwarf from the usual generalized uncertainty principle. Considering that the parameter signs of generalized uncertainty principles would change with different quantum gravity models, we provide two solutions to respectively restore the mass limit from two scenarios including positive and negative parameters, and the case of negative parameter is the focus of our research. For the case of negative parameter, a field-dependent parameter expression is derived, and it not only leads to a negative sign for white dwarf to restore its mass limit, but also leads to a positive sign to keep consistent with string theory and some models of quantum gravity. Finally, by combining with the solution for the case of positive parameter, we can see that the actual physical effect of the negative sign is aimed at limiting the exorbitant uncertainty of momentum. In addition, our paper also provides a feasible idea for further investigating the changes of parameter signs of generalized uncertainty principles in other areas of physics.

**References**

[1] R.H. Fowler, On dense matter, Monthly Notices of the Royal Astronomical Society, 87 (1926) 114-122.

[2] J.A. Wheeler, W.H. Zurek, Quantum Theory and Measurement, Philosophy of Science, 52 (1985).

[3] W. Pauli, Zur Quantenmechanik des magnetischen Elektrons, Zeitschrift für Physik, 43 (1927) 601-623.

[4] E.C. Stoner, LXXXVII. The equilibrium of dense stars, The London, Edinburgh, and Dublin Philosophical Magazine and Journal of Science, 9 (1930) 944-963.

[5] S. Chandrasekhar, The maximum mass of ideal white dwarfs, The Astrophysical Journal, 74 (1931) 81.

[6] A. Romero, S. Kepler, S. Joyce, G. Lauffer Ramos, A. Córsico, The white dwarf mass-radius relation and its dependence on the hydrogen envelope, Monthly Notices of

the Royal Astronomical Society, 484 (2019) 1-15.

[7] J.L. Provencal, H.L. Shipman, E. Hog, P. Thejll, Testing the White Dwarf Mass-Radius Relation withHipparcos, The Astrophysical Journal, 494 (1998) 759-767.

[8] V. Chandra, H.-C. Hwang, N.L. Zakamska, S. Cheng, A Gravitational Redshift Measurement of the White Dwarf Mass–Radius Relation, The Astrophysical Journal, 899 (2020) 146.

[9] J.D. Cummings, J.S. Kalirai, P.E. Tremblay, E. Ramirez-Ruiz, J. Choi, The White Dwarf Initial–Final Mass Relation for Progenitor Stars from 0.85 to 7.5 M ⊙, The Astrophysical Journal, 866 (2018) 21.

[10] M.R. Douglas, D. Kabat, P. Pouliot, S.H. Shenker, D-branes and short distances in string theory, Nuclear Physics B, 485 (1997) 85-127.

[11] F. Scardigli, Generalized uncertainty principle in quantum gravity from micro-black hole gedanken experiment, Physics Letters B, 452 (1999) 39-44.

[12] S. Hossenfelder, Minimal Length Scale Scenarios for Quantum Gravity, Living Reviews in Relativity, 16 (2013) 2.

[13] L.J. Garay, Quantum gravity and minimum length, International Journal of Modern Physics A, 10 (1995) 145-165.

[14] K. Konishi, G. Paffuti, P. Provero, Minimum physical length and the generalized uncertainty principle in string theory, Physics Letters B, 234 (1990) 276-284.

[15] M. Maggiore, A generalized uncertainty principle in quantum gravity, Physics Letters B, 304 (1993) 65-69.

[16] S. Capozziello, G. Lambiase, G. Scarpetta, Generalized Uncertainty Principle from QuantumGeometry, International Journal of Theoretical Physics, 39 (2000) 15-22.

[17] M. Faizal, Supersymmetry breaking as a new source for the generalized uncertainty principle, Physics Letters B, 757 (2016) 244-246.

[18] A. Sepehri, A. Pradhan, A. Beesham, On the origin of generalized uncertainty principle from compactified M5-brane, Modern Physics Letters A, 32 (2017) 1750123.

[19] A.N. Tawfik, A.M. Diab, A review of the generalized uncertainty principle, Reports on Progress in Physics, 78 (2015) 126001.

[20] R. Rashidi, Generalized uncertainty principle and the maximum mass of ideal

white dwarfs, Annals of Physics, 374 (2016) 434-443.

[21] F. Scardigli, The deformation parameter of the generalized uncertainty principle, Journal of Physics: Conference Series, 1275 (2019) 012004.

[22] L. Buoninfante, G.G. Luciano, L. Petruzziello, Generalized uncertainty principle and corpuscular gravity, The European Physical Journal C, 79 (2019) 663.

[23] R.J. Adler, P. Chen, D.I. Santiago, The Generalized Uncertainty Principle and Black Hole Remnants, General Relativity and Gravitation, 33 (2001) 2101-2108.

[24] Y.C. Ong, An effective black hole remnant via infinite evaporation time due to generalized uncertainty principle, Journal of High Energy Physics, 2018 (2018) 195.

[25] P. Pedram, K. Nozari, S.H. Taheri, The effects of minimal length and maximal momentum on the transition rate of ultra cold neutrons in gravitational field, Journal of High Energy Physics, 2011 (2011) 93.

[26] K. Nouicer, Quantum-corrected black hole thermodynamics to all orders in the Planck length, Physics Letters B, 646 (2007) 63-71.

[27] P. Pedram, A higher order GUP with minimal length uncertainty and maximal momentum, Physics Letters B, 714 (2012) 317-323.

[28] A. Mathew, M.K. Nandy, Existence of Chandrasekhar's limit in generalized uncertainty white dwarfs, Royal Society Open Science, 8  210301.

[29] Y.C. Ong, Generalized uncertainty principle, black holes, and white dwarfs: a tale of two infinities, Journal of Cosmology and Astroparticle Physics, 2018 (2018) 015.

[30] Y.C. Ong, Y. Yao, Generalized uncertainty principle and white dwarfs redux: How the cosmological constant protects the Chandrasekhar limit, Physical Review D, 98 (2018) 126018.

[31] J. Pinochet, M.V.S. Jan, Chandrasekhar limit: an elementary approach based on classical physics and quantum theory, Physics Education, 51 (2016) 035007.

[32] X.-D. Du, C.-Y. Long, The influence of approximation in generalized uncertainty principle on black hole evaporation, Journal of Cosmology and Astroparticle Physics, 2022 (2022) 031.

[33] H. Moradpour, A.H. Ziaie, S. Ghaffari, F. Feleppa, The generalized and extended uncertainty principles and their implications on the Jeans mass, Monthly Notices of the

Royal Astronomical Society: Letters, 488 (2019) L69-L74.

[34] U. Das, B. Mukhopadhyay, New Mass Limit for White Dwarfs: Super-Chandrasekhar Type Ia Supernova as a New Standard Candle, Physical Review Letters, 110 (2013) 071102.

[35] I. Hachisu, M. Kato, H. Saio, K.i. Nomoto, A single degenerate progenitor model for type Ia supernovae highly exceeding the Chandrasekhar mass limit, The Astrophysical Journal, 744 (2011) 69.

[36] S. Kalita, B. Mukhopadhyay, T.R. Govindarajan, Significantly super-Chandrasekhar mass-limit of white dwarfs in noncommutative geometry, International Journal of Modern Physics D, 30 (2021) 2150034.

[37] S. Kalita, T.R. Govindarajan, B. Mukhopadhyay, Super-Chandrasekhar limiting mass white dwarfs as emergent phenomena of noncommutative squashed fuzzy spheres, International Journal of Modern Physics D, 30 (2021) 2150101.

[38] J. Magueijo, L. Smolin, Lorentz Invariance with an Invariant Energy Scale, Physical Review Letters, 88 (2002) 190403.

[39] S. Chandrasekhar, On Stars, Their Evolution and Their Stability (Nobel Lecture), Angewandte Chemie International Edition in English, 23 (1984) 679-689.

[40] D. Gregoris, Y.C. Ong, On the Chadrasekhar Limit in Generalized Uncertainty Principles, arXiv preprint arXiv:2202.13904, (2022).

[41] F. Scardigli, R. Casadio, Gravitational tests of the generalized uncertainty principle, The European Physical Journal C, 75 (2015) 425.

[42] J. Zhang, Black hole quantum tunnelling and black hole entropy correction, Physics Letters B, 668 (2008) 353-356.

[43] S. Das, P. Majumdar, R. Bhaduri, General Logarithmic Corrections to Black Hole Entropy, Classical and Quantum Gravity, 19 (2001).

[44] S.M. Christensen, M.J. Duff, Axial and conformal anomalies for arbitrary spin in gravity and supergravity, Physics Letters B, 76 (1978) 571-574.

[45] I. Antoniadis, E. Gava, K.S. Narain, Moduli corrections to gravitational couplings from string loops, Physics Letters B, 283 (1992) 209-212.

[46] L. Xiang, X.Q. Wen, A heuristic analysis of black hole thermodynamics with

generalized uncertainty principle, Journal of High Energy Physics, 2009 (2009) 046-046.

[47] R. Banerjee, S.K. Modak, Exact differential and corrected area law for stationary black holes in tunneling method, Journal of High Energy Physics, 2009 (2009) 063-063.

[48] K. Nozari, S. Saghafi, Natural cutoffs and quantum tunneling from black hole horizon, Journal of High Energy Physics, 2012 (2012) 5.

[49] A. Kempf, G. Mangano, R.B. Mann, Hilbert space representation of the minimal length uncertainty relation, Physical Review D, 52 (1995) 1108-1118.

[50] A. Ghosh, P. Mitra, Counting of Black Hole Microstates, Indian Journal of Physics, 80 (2006).

[51] P. Bargueño, S. Bravo Medina, M. Nowakowski, D. Batic, Quantum-mechanical corrections to the Schwarzschild black-hole metric, EPL (Europhysics Letters), 117 (2017) 60006.

[52] R. Banerjee, B. Majhi, Quantum Tunneling Beyond Semiclassical Approximation, Journal of High Energy Physics, 2008 (2008).

[53] D. Fursaev, Temperature and entropy of a quantum black hole and conformal anomaly, Physical Review D, 51 (1995) R5352-R5355.

[54] R.J. Adler, D.I. Santiago, On gravity and the uncertainty principle, Modern Physics Letters A, 14 (1999) 1371-1381.

[55] C. Lunkes, Č. Brukner, V. Vedral, Equation of state for entanglement in a Fermi gas, Physical Review A, 71 (2005) 034309.

[56] F. Scardigli, G. Lambiase, E.C. Vagenas, GUP parameter from quantum corrections to the Newtonian potential, Physics Letters B, 767 (2017) 242-246.

[57] G.G. Luciano, L. Petruzziello, GUP parameter from maximal acceleration, The European Physical Journal C, 79 (2019) 283.

[58] T. Kanazawa, G. Lambiase, G. Vilasi, A. Yoshioka, Noncommutative Schwarzschild geometry and generalized uncertainty principle, The European Physical Journal C, 79 (2019) 95.

[59] E. Vagenas, S. Alsaleh, A. Farag Ali, GUP parameter and black hole temperature, EPL (Europhysics Letters), 120 (2018).

[60] P. Chen, Y.C. Ong, D.-h. Yeom, Generalized uncertainty principle: implications for black hole complementarity, Journal of High Energy Physics, 2014 (2014) 21.